# Research Proposal

**Novel sensing media based on ferromagnetic microwires for application to the remote imaging of the stress distribution and damage in the microwave range**

*Larissa Panina, Serghei Sandacci, and Dmitriy Makhnovskiy*[*)]

School of Computing, Communications and Electronics,
The University of Plymouth, Drake circus, Plymouth, Devon PL4 8AA, United Kingdom.

[*)]e-mails: dmakhnovskiy@plymouth.ac.uk    and    dmakhnovskiy@df.ru

**Summary**

The use of microwave for condition assessment of structural elements is becoming established as a non-destructive evaluation method in civil engineering, especially for detection of invisible damage inside structural bodies. This technology is based on a reconstruction of dielectric profile (image) of a structure illuminated with microwaves, through scattering measurement controlled by software. The controllable changes of the scattering response, and hence the damage detection are possible only for quite contrast variations in the material structure such as wide cracks or voids. Nevertheless, an interpretation of the microwave profile remains a very serious problem. Moreover, since a pre-damage stress does not result in any variations in the dielectric constant of usual structural materials, the excess stress and material fatigue become unpredictable. The mechanical stress is static in nature therefore an additional mediate physical process is required to visualise it.

*In this research project we propose a new composite medium, which can visualise the mechanical stress at any stage: before and after damage. The main feature of the proposed stress-tuneable composite is its permittivity (dielectric constant), which depends on the mechanical stress. This kind of composite material can be characterised as a "sensing medium" that opens up new possibilities for remote monitoring of stress with the use of microwave transceiving techniques. The composite material can be made as a bulk medium or as thin cover to image the mechanical stress distribution inside construction or on its surface.*

The composite consists of short pieces of ferromagnetic wires with *the helical magnetic anisotropy*, embedded into a dielectric matrix. In some cases the immediate construction material (such as concrete) can be used as the composite matrix. Also, the composite sample can be fabricated in the form of a thin layer of polymer matrixes with thickness less than 1 millimetre. The short wire inclusions play the role of "the elementary scatterers", when the electromagnetic wave irradiates the composite and induces an electrical dipole moment in each inclusion. These induced dipole moments form the dipole response of the composite, which can be characterised by some effective permittivity. The stress dependence of the effective permittivity of the composite arises from *the microwave magneto-impedance* of the ferromagnetic wire inclusions, which is highly sensitive to their internal strains. In the vicinity of the antenna resonance (related with the short wire inclusion) any variations in the magneto-impedance of wire result in large change of its dipole moment, and hence the effective permittivity. Therefore, this composite demonstrates both the stress tuneable and resonance properties (selective absorption). Stress-imaging can be achieved by microwave scanning the proposed media and analysing the reflected signal by continuous radar techniques.

**Objectives**

The overall objective of the project is to develop a sensing medium for the visualisation of mechanical stresses and the respective imaging methods. This approach is similar to the tomographic data acquisition method. The project includes:

1. The investigation of methods to induce helical magnetic anisotropy in amorphous microwires and the magnetic structures sensitivity to the external strains of various origins (substrates types, temperature, pressure and torque).
2. The investigation of microwave (1-40 GHz) magneto-impedance in amorphous microwires with helical anisotropy under the effect of applied stress.
3. The development of fabrication methods for producing the wire-based composites with different dielectric matrixes (concrete, polymers, epoxies, and glass) and the analysis of their effective electromagnetic functions by means of free-space measuring techniques.
4. The development of data recovery methods of internal stress distributions in the material using microwave reflection data.
5. Testing the proposed sensing media for such applications as remote safety monitoring of construction components (building wall and various industrial and transport corps).

**Beneficiaries**

There is a broad range of beneficiaries of the proposed research. Contemporary safety standards require high performance monitoring systems in many areas: civil constructions, airframe crack propagation, pipeline and pressure vessel structural condition. The proposed research is of wide interest in physics and engineering of new "metamaterials" with tuneable properties.

Over the years, we have close collaboration with leading manufacturers of amorphous wires: **MFTI SRL** (Republic of Moldova) and **UNITIKA LTD R&D** (Japan). There is also much interest from **Sensors Technology LTD** (UK), and **Transense Technologies PLC** (UK) in the commercial implications of this project.

## DESCRIPTION OF PROPOSED RESEACH PROJECT

**Background**

The increasing demands for improvements in safety for both industrial and civil applications require permanent monitoring of structural components. The use of traditional sensors and signal apparatus partly solves the safety requirements when only a small number of sensors are needed. For large objects, traditional methods are not practical and are therefore difficult to realise. Thus, to estimate the mechanical conditions of large surfaces or bulky constructions it is necessary to determine the mechanical stress distribution over a large number of regions, since cracks and excess stresses can not only occur in special critical areas, but also in arbitrary places with inherent structural imperfections. We propose that stress sensitive composites can be developed to enable remote monitoring of these stress distributions within the investigated material. The mechanical stresses are static in nature therefore we propose to utilise microwave radiation reflected from the special composite material where the microwave electromagnetic functions are strongly affected by external tensile stresses adjacent to the composite material.

The structure of the composite material with ferromagnetic wires embedded into the dielectric matrix is shown in Figure 1. The wire inclusions interact with the electromagnetic radiation similar to micro-antennas. At the antenna resonance, the dipole moment reaches its maximum value resulting in the most intensive scattering. Then, even small variations in the wire magneto-impedance caused by stress produce a considerable change in the current density distribution within the microwires and, consequently, in the induced dipole moment of the elementary wire-scatterer. Then, the resulting effect of stress will be a considerable change in the scattered wave. Therefore, the proposed composite demonstrates both the tuneable and resonance properties (selective absorption).

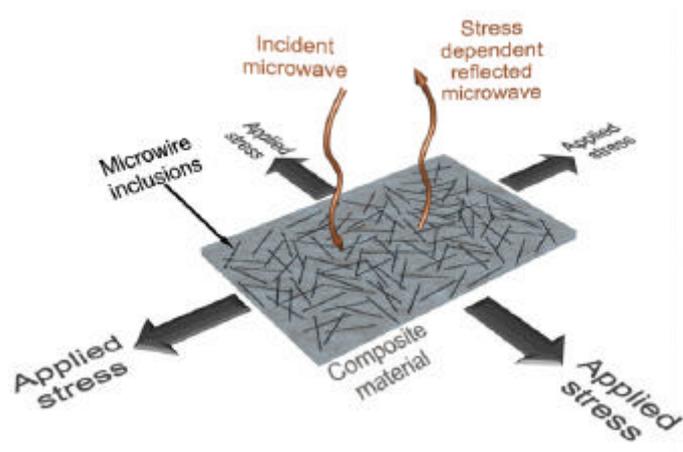

*Fig. 1 Composite material with ferromagnetic wire inclusions.*

Our research group at the University of Plymouth has started research activities on magnetic field tuneable composite materials consisting of ferromagnetic wires with circumferential anisotropy. However, such composite will be insensitive to the mechanical stress effects. *In this project we propose to utilise amorphous wires with helical anisotropy for the composite filling.* Then, the equilibrium magnetic structure of the wire and induced dipole moments will be very sensitive to the external stresses/torque. *Furthermore, the present proposal is aimed to develop the microwave free space technique to investigate stress-tuneable composite materials.*

**Ferromagnetic amorphous microwires**

Thin amorphous ferromagnetic wires are very promising for high frequency applications.[1-5] The alloy of composition $(Co_{1-x}Fe_x)_{75}Si_{15}B_{10}$ allows various magnetic structures to be realised. The sign and value of the magnetostriction, $\lambda$, which plays the main role to determine the magnetic behaviour (i.e. domain structure and hysteresis loop) depends on "x".[5,6] Positive ($x > 0.06$) and negative ($x < 0.06$) magnetostrictions result respectively in radial and circumferential easy axis in the shell, whereas the inner core always has the longitudinal magnetisation (although it can be very small). Therefore, the negative magnetostriction is typical for Co-rich alloys. For example, a wire of a composition $Co_{70.5}Fe_{4.5}Si_{15}B_{10}$ exhibits excellent soft magnetic properties having almost zero (but still negative) magnetostriction of $\lambda = -10^{-7}$.

Currently, there are two main techniques of wire fabrication. Amorphous wires are made by **UNITIKA LTD R&D** (Japan) using the in-water spinning method, and then cold drawn from a diameter of about 125 μm of the as-cast wire to diameters of 30 μm.[1,2] The final sample undergoes annealing with a tension stress to build up a certain magnetic structure. This method requires very careful control of the annealing process to obtain repeatable magnetic parameters. Some commercial companies (for example, **MFTI SRL**, Republic of Moldova) and research laboratories produce amorphous wires with a glass coating, by a modified Taylor-Ulitovskiy casting method.[7] The metallic core (ranging between 5-50 microns) has an amorphous and/or microcrystalline microstructure in order to achieve the desired magnetic properties, such as magnetic anisotropy and coercivity. The ratio of the metal core and glass coating thickness also affects the magnetic properties.[6] The fabrication method of glass-covered microwires introduces a quite large internal stress mainly arising from the difference in thermal expansion coefficients of metal (nucleus) and glass (sheath). The value of this internal stress, which has the tensor nature, can be easily controlled by the ratio of diameters of the glass cover and the metal core.

## Giant magneto-impedance (GMI) effect

The stress dependence of the reflected radiation from the proposed wire-based composite materials arises from microwave magneto-impedance of a ferromagnetic wire, which is highly sensitive to the internal strains. In simple terms, the giant magneto-impedance effect (discovered in around 1994 [8,9]) is understood as a colossal change of the complex resistance (impedance) of a ferromagnetic sample subjected to a high frequency current and a dc external magnetic field. The impedance variations are determined by the so-called magnetic skin-depth depending upon the ac permeability and equilibrium magnetic structure of the wire.[10] In fact, this structure can be modified not only by a dc magnetic field, but also by other external factors such, for example, as stress.[11-17] The stress dependence of the impedance became the reason to distinguish between the magneto-impedance (effect of a dc magnetic field) and stress-impedance (effect of stress via reverse magnetostrictive effect) effects. Nevertheless, in both cases the impedance changes are originated by the modification in the magnetic structure.

Up to the present, the GMI effect has been investigated for use in highly sensitive magnetic or stress sensors at megahertz frequencies.[2,18-21] However, it has been shown [22,23] that the field or stress sensitivity of the impedance in wires with circumferential and helical anisotropies remains very high even at the GHz range. This property can be used in microwave materials where the electromagnetic functions (permittivity and permeability) become sensitive to the mechanical or field actions.[24]

## Stress dependence of the magneto-impedance in the microwave range

The proposed media makes use of microwave stress-impedance. This effect in Co-rich amorphous wires was investigated over the MHz frequency range and demonstrated very high stress sensitivity (with a gauge factor about 2000). It was then proposed to utilise this effect in the development of stress sensors.[11-17] The sensor operation is based on a GMI wire incorporated into a self-oscillating circuit [18-21] where the voltage measured across the wire is related to the applied tensile stress or torsion. For remote stress sensing, the measuring technique is quite different and is based on the detection of the reflected radiation, which is a function of the stress-impedance.

To provide the strong stress sensitivity of GMI, the wires with a helical anisotropy have to be used. In the MHz range, the ac permeability of wires with the circumferential or helical anisotropies depends strongly on the effective anisotropy field, which is utilised in stress- and magneto-impedance sensors. This ac permeability can be changed by the dc external magnetic field or applied stress. However, with increasing frequencies, this dependence becomes weaker and weaker and almost disappears in the microwave region.[22-24] Then, the high frequency impedance will depend only on the static magnetisation orientation, which also can be changed by the field or stress.[24] Figure 2(a) demonstrates typical *"valve-like" behaviour* of the magneto-impedance that was measured in a wire with the circumferential anisotropy.[22,23] The impedance switches from one stable level to the other, following the dc magnetisation shown in Figure 2(b). Typical magneto-impedance characteristics with two peaks at the anisotropy field $H_K$ are seen for MHz frequencies (Fig. 2(a)). With increasing frequency, the low-field region ($H_{ex} < H_K$) in the impedance plots preserves its high sensitivity although the impedance maximum value at $H_{ex} = H_K$ decreases substantially. The transformations in the high-field ($H_{ex} > H_K$) region are more dramatic. With increasing frequency the impedance looses its sensitivity for higher fields and at frequencies above 1.6 GHz it remains constant for any $H_{ex} > H_K$. Therefore, the low level of the impedance corresponds to the circumferential direction of the magnetisation whereas the high level – to the longitudinal one (saturation).

*We predict similar valve-behaviour in the GHz range for wires with a helical anisotropy, where the magnetisation direction can be changed not only by the dc external magnetic field but also the mechanical stress.* For the valve stress impedance we should obtain similar curves as in Figure 2(a), where the field has to be substituted for the magnetisation angle which is proportional to the stress.

The tensile stress induces the magnetic anisotropy along the stress in the case of a positive magnetostriction ($\lambda>0$) and in the perpendicular direction for $\lambda<0$. The induced anisotropy defines the preferred direction of the static magnetisation. At zero dc external magnetic field, the tensile stress in both cases of $\lambda>0$ and $\lambda<0$ does not produce any change in the equilibrium magnetisation reinforcing only the effective anisotropy field: the circumferential for $\lambda<0$ and the longitudinal for $\lambda>0$. Only in the case of a helical anisotropy, when the equilibrium magnetisation is deflected from the circumferential or longitudinal directions (at zero magnetic field), the magnetisation rotation due to the tensile stress becomes possible. Therefore, in the GHz range only wires with a helical anisotropy can provide the strong stress dependence of the magneto-impedance.

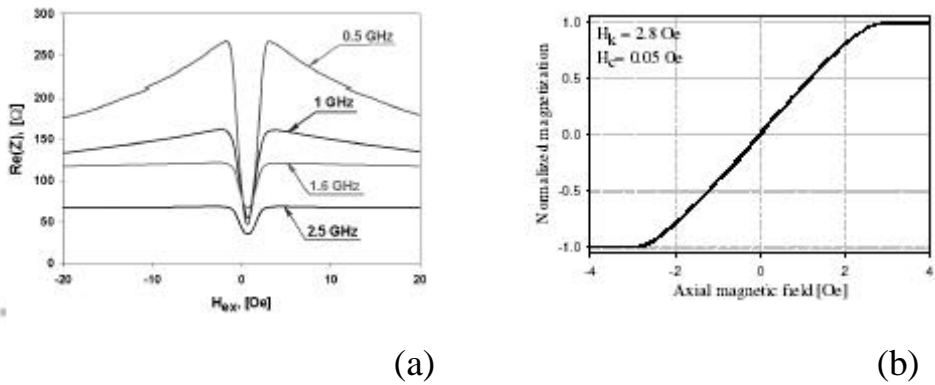

(a)          (b)

*Fig. 2 Typical "valve-like" behaviour of the magneto-impedance measured in a wire with the circumferential anisotropy in (a), when it switches from one stable level to the other, following the dc magnetisation in (b). The normalised magnetisation means the longitudinal projection with respect to the wire axis. $H_K = 2.8\ Oe$ is the anisotropy field and $H_c = 0.05\ Oe$ is coercivity.* (**After S. I. Sandacci et al, [22]**)

In order the applied stress could change the static magnetisation, the initial residual stress distribution in wires has to differ from the tensile stress. One possibility is to realise somehow a "frozen" torsion stress and corresponding helical anisotropy. This presents a rather complicated technological problem, especially, in wires with diameters less than 15 μm that are required for microwave applications.

An "artificial" helical magnetisation can be achieved in wires with a circumferential anisotropy ($\lambda<0$) by the combination of the tensile stress and the fixed dc longitudinal magnetic field $H_{ex}$, which should be chosen about of the anisotropy field $H_K$. The tensile stress reinforces the circumferential anisotropy, whereas $H_{ex}$ deflects the magnetisation from the initial circumferential direction. These two factors allow the magnetisation rotation due to the tensile stress at the fixed $H_{ex} \sim H_K$. In turn, this will result in the stress sensitivity of the magneto-impedance in the GHz range. Thus, using the wires with a circumferential anisotropy as the filling inclusions in the composite material, we obtain the "double-tuneable" composite where the stress sensitivity can be controlled by the dc external magnetic field.

Tension-annealed CoSiB amorphous wires of 30 **m**m in diameter (magnetostriction $l = -3 \cdot 10^{-6}$) posses a spontaneous helical anisotropy due to a residual stress distribution, and a relatively large anisotropy field.[25] In other cases, annealing under a torsion stress or current annealing in the presence of the axial magnetic field can induce a helical anisotropy.[17,26] In all the cases, the anisotropy angle is difficult to control. Therefore, for practical application the reliable method of inducing the helical anisotropy must be developed. *Some promising techniques include laser annealing [27] and anisotropic layering,[28] which are intended to be developed here.*

**Programme and Methodology**

The work is subdivided into 4 main parts **A**-**D**. It starts (**A**) with the investigation of technological aspects of inducing helical anisotropy in amorphous wires and their static magnetisation processes as a function of the applied stress. In (**B**), we will investigate the microwave properties of the individual wires incorporated into various materials by measuring its impedance with non-contact methods. The next stage (**C**) is the development of the technology to fabricate wire-based composite samples with different dielectric matrixes for the investigation of their effective permittivity and permeability by means of free space technique. An important extension to this investigation is to carry out measurements of the composite materials in actual real application environments (**D**) when the dielectric matrix is represented by immediate construction material or a layered composite material is attached to the construction surface. Finally, on the basis of the obtained results (**A**-**D**) the algorithm for reproducing the distribution of the mechanical stress will be developed. Therefore, the project is carefully integrated combining technological, physical and engineering problems. The integration between different parts can be clearly seen from the diagrammatic work plan.

*Stage A*

*A.1* The important initial research will be to establish reliable methods of inducing helical anisotropy in amorphous wires. Traditional methods such as annealing under torsion and even current annealing in an axial magnetic field do not guarantee reproducible anisotropy angles, which is of great importance for stress dependencies.[11-17,26] These methods may require further modifications. Along with this, new techniques such as laser annealing and layering with combined sputtering and electroplating are reported to produce glass-covered microwires with controllable magnetic properties. These techniques can also be used to induce a helical anisotropy with asymmetrical treatment. Another promising possibility is inducing helical anisotropy during the microwire fabrication. This will be done in collaboration with MFTI SRL who will provide glass-coated wires for this project. We also have collaboration with Unitika LTD and Tamag LTD (Spain) who also produce amorphous microwires

*Milestone 1* Establishment of reliable treatment to induce a helical anisotropy with a predetermined angle.

*A.2* The static magnetisation process (B-H loops) will be measured under the effect of various applied stresses to find the optimal stress-sensitivity with controllable technological treatment.

*Milestone 2* Sensitive control of DC magnetisation loops in wires with induced helical anisotropy by applied stresses.

*Stage B*

At this stage the microwave impedance of an individual wire will be investigated by non-contact method when the wire peace is placed on the top of the matched stripe line. In such configuration, the polarisation of the excitation field is close to that which exists in the plane wave: longitudinal electrical and transverse magnetic fields. Therefore, the signal reflected from the stripe-cell will contain information about scattering and resistive losses into the wire.

***B.1*** Development of the measuring technique of the wire microwave impedance by non-contact method allowing the application of external tensile stress

***Milestone 3*** Experimental setup for non-contact measurement of the wire impedance at microwave range (1-40 GHz).

***B.2*** Investigation of the applied stress effect on the wire impedance. Investigating the effect of various substrates (glass, ceramic, polymers) to which a wire is attached. Comparison with the results of the stress effect on the dc magnetisation loops. Modelling the obtained data by calculating surface impedance in a wire with a helical anisotropy that is controlled by the applied stress.

***Milestone 4***. Building database representing microwave impedance as a function of a tensile stress in various wire materials.

## *Stage C*

Stress-impedance effect in an individual wire constitutes the microwave response from wire-assembly incorporated into a certain dielectric matrix. This will require firstly the investigation into the technology of wire-based composite materials. Two kinds of materials with random and ordered structures will be considered as shown in Figure 2(a) and 2(b), respectively. The microwave properties of wire-based composites in the form of a layered material will be investigated by free-space technique.

***C.1*** Fabrication of composite materials with inclusions of wire. We are aimed to the investigation of the resonance properties of the effective electromagnetic functions, which result from the antenna resonances at individual wires. Therefore, the wire length and the permittivity of the matrix will determine the resonance frequency. Two composite structures will be investigated:
*(1) Ordered wires on/in dielectric substrate.* Such kind of structures have strongly anisotropic microwave response with respect to the polarisation of the incident electromagnetic wave. This anisotropy introduces an additional parameter, which may be useful for the selective properties of the composite structure.
*(2) Randomly oriented wires mixed into a polymer layer.* The initial experiments will combine epoxy resin and microwave ceramic powder with ferromagnetic wire inclusions. The microwave ceramic powders (for example, provided by **Morgan Electro Ceramics of Wrexham, UK**) have dielectric constants ranging between 10 and 80 with particle sizes of approximately 1 micron and are very useful to control the resonance frequency.

***Milestone 5*** Samples of composite materials with microwire inclusions.

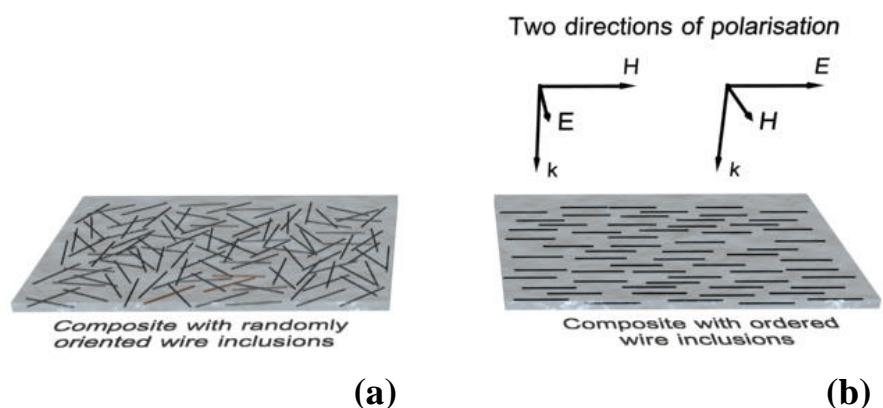

*Fig. 3 Two types of composites with randomly oriented wire inclusions in (a) and ordered wire inclusions in (b).*

*C.2* The effective permittivity and permeability of wire-based composite materials will be investigated by the microwave free-space measurement technique (MFSMT).[29-34] The reflection and transmission coefficients are the measured parameters. In the case of proposed composite material, these parameters are related to internal stress in the material. The main advantage of this technique is that with suitable modifications, it is possible to make precise, accurate and reproducible remote measurements on composite materials under natural environmental conditions. The spatial resolution of MFSMT depends on the wavelength of the electromagnetic wave. For the microwave band of 1-40 GHz, wavelength varies from 300 mm to 7 mm. But, modern electronics and computer processing will improve its potential for industrial applications. Since the penetration of microwaves in good conducting materials is very small, MFSMT is mainly used for nonmetallic materials. Our composite material has an effective dielectric permittivity with the Lorentz dispersion, which demonstrates resonance absorption ("artificial dielectric" with losses).[35,36]

Figure 4 demonstrates typical resonance dispersion curves (of the Lorentz type) of the effective permittivity $\varepsilon_{eff}(\omega)$ calculated at the GHz range for two magnetisation directions in the wire inclusions: (i) circumferential magnetisation and (ii) near longitudinal magnetisation.[24] The matrix permittivity (composite host) is $\varepsilon = 16$ and the volume concentration of inclusions is $p = 0.01\%$. This concentration is considerably smaller than the percolation threshold $p_c \propto (2a/l) \times 100\% \sim 0.1\%$ ($2a = 10 \mu m$ and $l = 1 cm$). The effect of the magnetisation direction shows up in changing the character of the dispersion curves. For the circumferential direction (low impedance) the dispersion curves are of a resonance type: at $f = f_{res}$ the imaginary part reaches a maximum and the real part equals zero. For near longitudinal direction (high impedance), the impedance is increased and, as a consequence, the internal losses in the inclusion, which results in the dispersion of a relaxation type. In work [36] the transformation of the dispersion $\varepsilon_{eff}(\omega)$ from a resonant type to relaxation one was associated with a different wire conductivity, which defines the resistive loss. In our case it is provided through the stress- or field- dependent impedance instead of conductivity. At the wire concentration larger than a certain value (for example, $p = 0.01\%$) the real part of $\varepsilon_{eff}(\omega)$ becomes negative in the vicinity of the resonance. Changing the magnetisation direction (by the stress or field), it is possible change gradually Re($\varepsilon_{eff}(\omega)$) from negative to positive values.

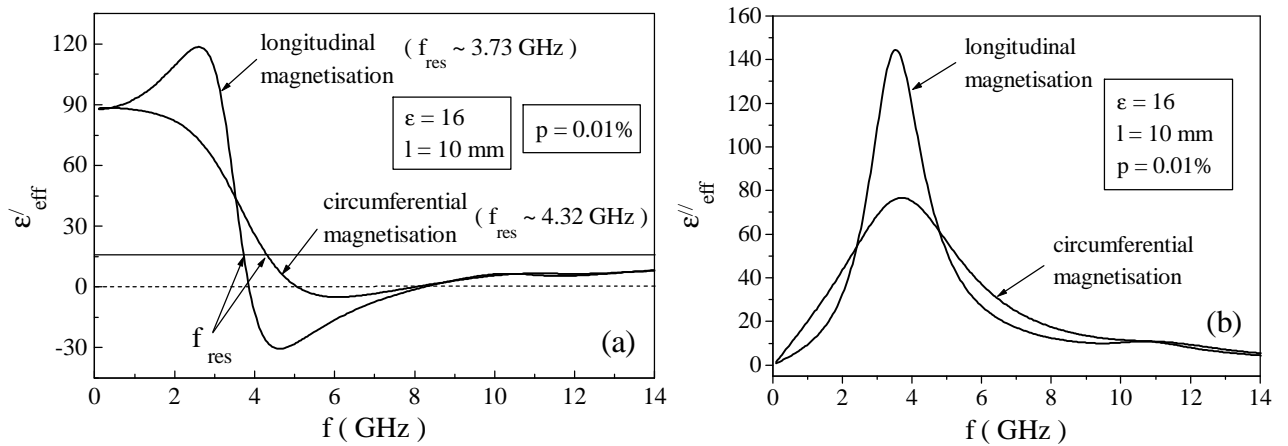

***Fig. 4*** *Transformation of the dispersion of the effective permittivity $\varepsilon_{eff} = \varepsilon' + i\varepsilon''$ from a resonance type to a relaxation one due to the magnetisation direction in the vicinity of the antenna resonance. The inclusion concentration is $p = 0.01\%$ and matrix permittivity is $\varepsilon = 16$. The inclusion length is 1 cm and its diameter is 10* **mm**. *(After D. P. Makhnovskiy and L. V. Panina [24])*

Before we discussed the bulk dielectric properties of the composite, which can be characterised by an effective permittivity $\varepsilon_{eff}$. If the composite sample is fabricated in the form a thin slab, the effective permittivity takes the matrix form:[24]

$$\hat{\mathbf{e}}_{eff} = \begin{pmatrix} \varepsilon_{eff} & 0 & 0 \\ 0 & \varepsilon_{eff} & 0 \\ 0 & 0 & \varepsilon \end{pmatrix},$$

where $\varepsilon_{11} = \varepsilon_{22} = \varepsilon_{eff}$ are the "tangential" effective permittivities in the plane of the composite sample if the wire inclusions have random orientations ($\varepsilon_{11} \neq \varepsilon_{22}$ if the wire inclusions are ordered, as shown in Fig. 3(b)), and $\varepsilon_{33} \approx \varepsilon$ is the "transverse" permittivity with respect to the composite surface.

The following four classes of the microwave free-space measurement techniques can be used for the investigation of the effective permittivity:

*1) Free-space technique operating in the far-field region employing transceiving antennas.[29,30]*

*2) Free-space technique operating in near-field region, which employs a metal, shut behind the composite slab, which is placed in close contact to the transceiving horn antenna.[35]*

*3) Free-space technique using the microwave reflectometer.[30]*

*4) Waveguide technique operating in the near-field region which employs open-ended coaxial line, rectangular waveguide, microstrip line and cavity resonator as probes.[29]*

Microwave nondestructive testing methods are fast, accurate and continuous techniques for evaluation of the material structure and its effective parameters. In general, these techniques involve measurement of reflection and transmission. Figure 5(a) gives an example of the free-space microwave measurement system consisting of a pair of horn antennas with spot-focusing polymer lenses. One lens gives an electromagnetic plane wave and the other one focuses the electromagnetic radiation passed through the sample. This system allows the investigation of both the reflection and transmission properties. The free-space method for near-field measurements, shown in Figure 5(b), employs a metal shut behind the composite slab, which is placed in close contact to the transceiving horn antenna.

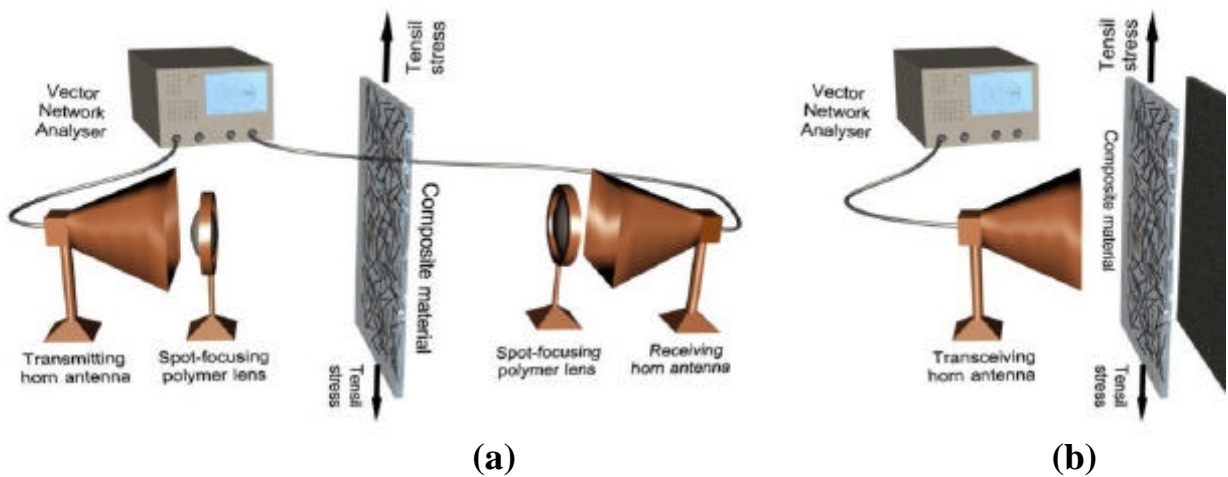

*Fig. 5 Schematic diagrams for microwave free-space measurements in the far-field region in (a) and the near-field region in (b).*

*The free-space technique operating in the far-field region is the most appropriate for the proposed research since it is contactless and close to the real condition of the remote monitoring.* This method has been employed for detecting voids and debonding between the polymer jacket and the concrete column, which may significantly weaken the structural performance of the column otherwise attainable by jacketing.[30] This so-called jacketing technology using fiber reinforced polymer (FRP) composites is being applied for seismic retrofit of reinforced concrete (RC) columns designed and constructed under older specifications. The microwave imaging technology is based on the reflection analysis of a continuous electromagnetic wave (EM) sent toward and reflected from layered FRP-adhesive-concrete medium. A poor bonding conditions including voids and debonding will generate air gaps which produce additional reflections of the EM wave. The dielectric properties of various materials involved in the FRP-jacketed RC column were first measured. Then, the measured properties were used for a computer simulation of the proposed EM imaging technology. To increase the method resolution, the use of the dielectric lenses is invoked to focus the electromagnetic wave on the interested region. Waves reflected from the other regions where the beam is defocused will have relatively small amplitude, and thus the differences between perfect bonding and imperfect one can be detected more effectively in this way. The FRP-jacketed concrete columns under testing are illustrated in Figure 6(a) and 6(b) for the different types of focusing system: (a) transmitting and receiving dielectric "focusing cones" and (b) transceiving horn antenna with a dielectric lens.

*The microwave imaging technology developed in work [30] can be improved and adopted for our applications. Most probably, the focusing system will not be required to visualise the stress distribution since a non-diffraction mechanism is involved. Therefore, the distance between the transceiving antenna and the measured structure can be significantly increased. In this case, a radar scanning of a big construction area becomes possible.*

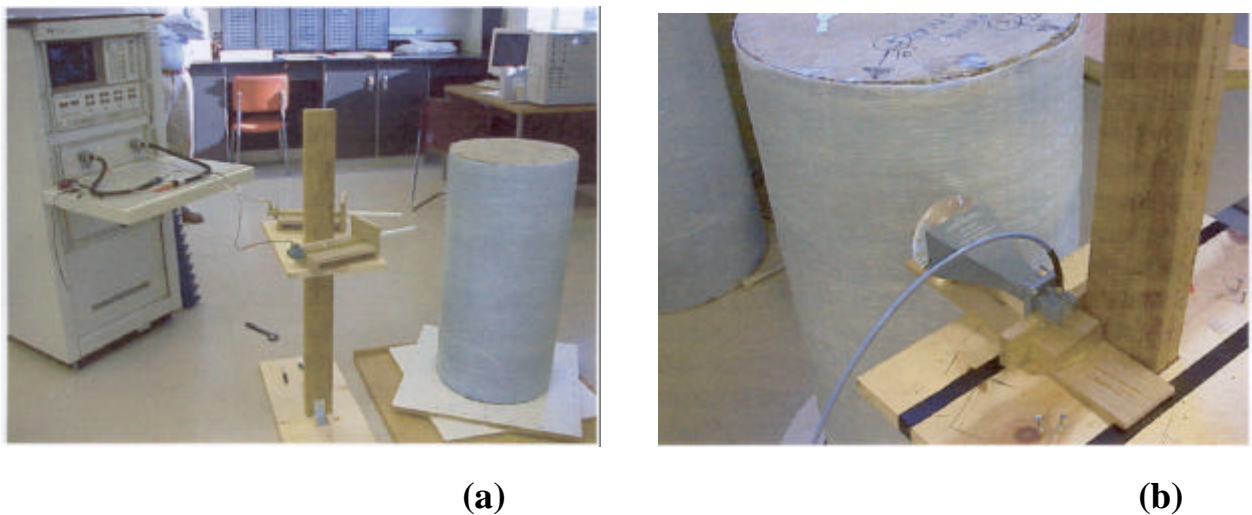

**(a)**          **(b)**

***Fig. 6*** *The FRP-jacketed concrete columns under testing for the different types of focusing system: (a) transmitting and receiving dielectric "focusing cones" and (b) transceiving horn antenna with a dielectric lens.* (***After M. Q. Feng et al, [30]***)

The free-space techniques operating in near-field region require the composite material to be in close contact with the probe. Therefore, they are not precisely contactless (the measurements are made in near-field region). *Nevertheless, these methods may appear more useful to evaluate the effective permittivity of the composite.* There are usually two fundamental difficulties associated with the measurement of the free space reflection and transmission with respect to the material sample: first, it is difficult to generate a uniform plane wave and second, unwanted parasitic reflections and multiple reflections are produced in the measurement path. The microwave reflectometer used in work [30] eliminates both of these problems. *This method looks most perspective for evaluating the effective permittivity of the composite.* The design of the reflectometer is based on the use of hollow metal dielectric waveguide (HMDW) as the EM wave transmission line structure or medium. The HMDW consists of a large (compared to wavelength) rectangular waveguide with thin non-resonant dielectric and absorber layers on two opposite walls, or on all four walls. The aperture of the reflectometer is of the order of 7 wavelengths by 7 wavelengths square. As shown in Figure 7, the reflectometer has four ports.

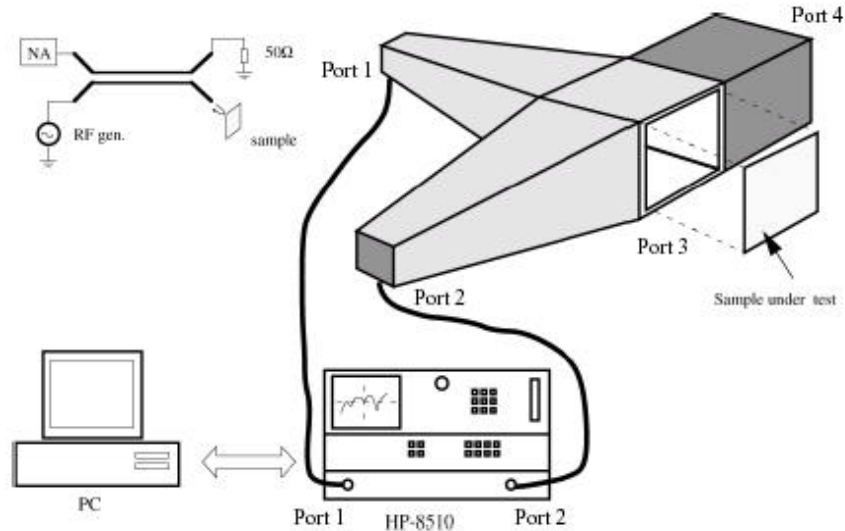

*Fig. 7 Experimental setup for dielectric property measurements (**After M. Q. Feng et al, [30]**)*

A microwave signal from port 1 of the Network Analiser is fed into the small end of the tapered transition connected to port 1 of the HMDW cross directional coupler. The signal passes through the HMDW cross waveguide coupler at port 1. Part of the signal goes to the load at port 4, while most of the signal passed through, to the composite sample under test at port 3. The reflection from the sample returns through port 3 and in turn splits into a component that passes through to the source arm at port 1 and into a major component that is coupled to the receiver/detector arm at port 2 of the HMDW coupler. The coupled signal passes through the waveguide transition of arm 2 and onto port 2 of the Network Analiser. The magnitude of the received signal at port 2 of the Network Analiser is proportional to the magnitude of the reflection coefficient of the material sample under test.

*Milestone 6* Setup and calibration of microwave free space measurement techniques.

*Milestone 7* Experimental investigations of the composites microwave properties.

### *Stage D*

The research of stage C will be followed by the investigation of the wire-composite materials in natural environment by microwave scanning techniques, which are similar to far-field MFSMT. Firstly, the condition of layered composite material attached to the construction surface will be investigated. The change in the stress distribution will be decided by analysing contrast mapping. This stage presents an advanced research and at present time can be foreseen in general terms. It is expected that experience in data interpretation obtained in computer tomography can be helpful in this analysis. We propose to use calibrated microwave labels insensitive to the stress, for example, materials with non-magnetic wires, which can provide some reference signals. Final stage would be to use real construction materials with incorporated microwires. This stage will require collaboration with civil engineering material research.

*Milestone 8* Mapping of stress distribution in interrogated objects.

**Relevance to beneficiaries**

There is a broad range of beneficiaries of the proposed research. Contemporary safety standards require high performance monitoring systems in many areas: civil constructions, airframe crack propagation, pipeline and pressure vessel structural condition. The proposed research is of wide interest in physics and engineering of new "metamaterials" with tuneable properties.[37,38] Most interesting applications of proposed novel type of stress-sensitive composites are:
- *Monitoring (diagnostic and prognostic) of bridges and building components, oil tanker corps, airframe cracks propagation, and pipeline and pressure vessel structural condition.*
- *Tyre pressure monitoring system.*